\documentclass[aps,reprint,superscriptaddress]{revtex4-1}

\usepackage{amssymb}
\usepackage{amsmath}
\usepackage{amscd}
\usepackage{latexsym}
\usepackage{graphicx}

\usepackage{enumitem}

\newcommand{\be}{\begin{equation}}
\newcommand{\ee}{\end{equation}}
\newcommand{\Dlt}{\Delta}

\newcommand{\prt}{\partial}
\newcommand{\br}{{\bf r}}
\newcommand{\bd}{{\bf d}}
\newcommand{\bfe}{{\bf e}}
\newcommand{\bn}{{\bf n}}
\newcommand{\bP}{{\bf P}}

\newcommand{\bE}{{\bf E}}
\newcommand{\bS}{{\bf S}}

\newcommand{\al}{\alpha}
\newcommand{\sgm}{\sigma}

\newcommand{\gm}{\gamma}
\newcommand{\om}{\omega}
\newcommand{\Om}{\Omega}

\newcommand{\ra}{\rightarrow}

\newcommand{\lbd}{\lambda}

\newcommand{\rgl}{\rangle}
\newcommand{\lgl}{\langle}

\begin{document}

\title{Ultrafast polarization switching in ferroelectrics}

\author{V.I. Yukalov}

\affiliation{Bogolubov Laboratory of Theoretical Physics,
Joint Institute for Nuclear Research, Dubna 141980, Russia}

\affiliation{
Instituto de Fisica de S\~{a}o Carlos, Universidade de S\~{a}o Paulo,
CP 369, S\~{a}o Carlos 13560-970, S\~{a}o Paulo, Brazil}

\author{E.P. Yukalova}

\affiliation{Laboratory of Information Technologies,
Joint Institute for Nuclear Research, Dubna 141980, Russia}

\begin{abstract}   
A method of ultrafast switching of ferroelectric polarization is suggested. 
The method is based on the interaction of a ferroelectric sample with the feedback 
field of a resonator in which the sample is inserted. The polarization reversal
time can be of order of femtoseconds. The polarization switching produces a 
coherent electromagnetic pulse.  
\end{abstract}

\maketitle

\section{Introduction}

Ferroelectric materials, possessing spontaneous electric polarization, can be harnessed 
for various electronic devices \cite{Uchino_1,Dawber_2}. For example, they are used in 
devices regulating tunneling resistance \cite{Dawber_2} and enabling nonvolatile memory
\cite{Araujo_3}, in memristors \cite{Chanthbouala_4}, in neuromorphic networks 
\cite{Jo_5,Kim_6}, and in solar cells \cite{Zhou_7,Li_8}. 

To regulate the processing of such devices, it is often necessary to be able to quickly 
vary the direction and magnitude of ferroelectric polarization. There exist two ways of
polarization switching that can be called inhomogeneous (or incoherent) and homogeneous
(or coherent). 

First, the inhomogeneous way of polarization switching has been studied, being realized 
through the nucleation and growth of domains of opposite polarization, with moving domain 
walls under the influence of static electric fields 
\cite{Kolmogorov_48,Avrami_49,Merz_9,Ishibashi_50,Tagantsev_51}. This way, however, provides 
rather long switching times, on the order of nanoseconds, being limited by the domain 
recrystallization time that is typically hundreds of picoseconds 
\cite{Uchino_1,Dawber_2,Merz_9,Spierings_10,Xu_10}. Similar slow switching in the nanoscale 
volume of a ferroelectric can be realized by mechanical deformation of a ferroelectric 
sample \cite{Lu_10}. Another slow mechanism of polarization switching is due to the chemical 
oxidization at the surface of a ferroelectric film \cite{Wang_10}. Because of the principal 
restriction of the inhomogeneous switching caused by the limited domain recrystallization 
time, it has been necessary to find other ways that could provide much faster switching.   

The other way that has been developed relatively recently is the homogeneous (or coherent) 
switching process realized by external alternating fields, applied perpendicular to the 
ferroelectric polarization, in the optical \cite{Fahy_11,Ciaran_52,Li_12,Istomin_53},  
terahertz \cite{Takahashi_13,Cavalleri_14,Qi_15,Katayama_16,Chen_54,Hauf_17,Hoegen_18}, or
infrared \cite{Subedi_19} regions. Under this process, the alternating field acts directly 
on all ions of a single-domain sample, and polarization switching occurs through a 
continuous homogeneous mechanism, without formation of new domains of opposite polarization. 
Homogeneous switching is facilitated in films of nanometer thickness, where inhomogeneous 
nucleation is strongly suppressed \cite{Highland_55}. 
 
In order to realize a homogeneous switching, it is necessary that, first, all characteristic 
times of the process be much shorter than the domain nucleation-growth time and, second, 
that the sample be a single-domain ferroelectric. Under the ultrafast switching by means 
of alternating fields, the first condition is easy to accomplish, since the domain 
nucleation-growth time is sufficiently long, being of order of nanoseconds. And the 
preparation of single-domain ferroelectrics is a technical problem having several solutions 
\cite{Bratkovsky_56,Aguado_57,Bratkovsky_58,Pertsev_59}. For instance, single-domain states 
can be made stable by using strain \cite{Pertsev_59} or doping with point defects 
\cite{Zhang_60}. Also, there are plenty of ferroelectric films of nanometer thickness, 
where domain nucleation is suppressed \cite{Highland_55}. 

The homogeneous switching mechanism under the action of an alternating field, involving no 
nucleation and growth of oppositely polarized domains, can provide reversal times of order 
of picoseconds. However the present sources do not provide the strength of a pulse 
sufficient for completely switching the polarization. Experiments \cite{Mankowsky_20} have 
shown that the reversal can be only $40 \%$ of its equilibrium value. Although the reversal 
is quite fast, occurring in about $10^{-13}$ s, but the reversed polarization rapidly, 
during the same $10^{-13}$ s, returns to the initial state, similarly to the dynamics 
induced by terahertz pulses, when the reversal happens over a picosecond time scale, 
followed by its fast complete retrieval \cite{Rana_21}. 

Finally, time-dependent density functional theory simulations show that by strongly 
exciting electrons via laser pulses it could be possible to change the underlying 
dynamical potential energy surface, which could result in the polarization switching 
within tens of picoseconds \cite{Lian_22}.

In the present paper, we consider the homogeneous way of polarization switching involving
alternating fields, but with a rather different setup. The idea of the method is to put a 
ferroelectric, subject to an external constant electric field, into a resonator cavity. 
Then the polarization motion induces a resonator feedback field acting back on the 
ferroelectric. In such a way, there is no need in additionally imposed external 
electromagnetic pulses, but the ferroelectric produces the required pulse by itself through 
the feedback field. The polarization switching can be realized in femtoseconds. The 
suggested method of switching also uses alternating fields, similar to the techniques 
employing oscillating fields with fixed properties. However the principal difference is 
that here the alternating field is not imposed by external sources, but is self-organized, 
being created by moving polarization itself. Such a self-organized feedback field turns out 
to be essentially more effective than an externally imposed field.    

We use the system of units where the Planck constant $\hbar$ is set to one.

\section{Evolution equations}

We consider a ferroelectric inserted into a cavity. Generally, if the sample is sufficiently 
large and especially when it is in contact with other media, say dielectric, then it can be 
separated into domains \cite{Levanyuk_47}. But we consider the case of a cavity 
containing no other materials inside it, except the ferroelectric itself, which is a 
single-domain sample. 

Our main aim is to illustrate the idea of using a self-organized field of a resonator for 
accelerating polarization switching. We do not claim to treat a specific material, but we 
demonstrate the efficiency of the idea by a model. For this purpose, let us take the 
Hamiltonian in the pseudospin representation \cite{Blinc_23,Blinc_24}, having the form of
an Ising-type model in a transverse field. 
\be
\label{1}
 \hat H = - \Om \sum_j S_j^x  \; - \; \frac{1}{2} \sum_{i\neq j} J_{ij} S_i^z S_j^z \;
- \; \sum_j \bE_{tot} \cdot \bP_j \;  .
\ee
Here $S_j^\alpha$ is an $\alpha$-component of the $S = 1/2$ spin operator characterizing 
an electric dipole at site $j$, $\Om$ is tunneling frequency, $J_{ij} = J_{ji} > 0$ 
describes the strength of dipolar interactions, ${\bf E}_{tot}$ is the total electric 
field acting on dipoles, and 
\be
\label{2}
\bP_j = d_0 \bS_j
\ee
is a dipolar operator. 

In what follows, the total electric field $\bE_{tot}$ will have two components that can 
be called longitudinal and transverse. It is important that the polarization would also 
have these two nonzero components. 

This form of the Hamiltonian provides a good description of the so-called 
order-disorder ferroelectrics, although it can also be a reasonable approximation for 
other types of ferroelectrics \cite{Blinc_23,Blinc_24}. Among order-disorder 
ferroelectrics, it is possible to mention such as
$$
{\rm KH}_2{\rm PO}_4 \; , ~~ {\rm KH}_2{\rm As O}_4 \;, ~~ 
{\rm RbH}_2{\rm PO}_4 \; , ~~ {\rm RbH}_2{\rm As O}_4 \; ,
$$
$$
{\rm CsH}_2{\rm PO}_4 \; , ~~ {\rm CsH}_2{\rm As O}_4 \;, ~~
{\rm NH}_4{\rm H}_2{\rm PO}_4 \; , ~~ {\rm NH}_4{\rm H}_2{\rm As O}_4 \; ,
$$
and their deuterated analogs, in which ${\rm H}_2$ is replaced by ${\rm D}_2$. Similar 
Hamiltonians also are used for describing relaxor ferroelectrics \cite{Bokov_25}.  

Note that usually in the case of order-disorder ferroelectrics with spatially symmetric 
double wells, the vector of polarization possesses only the longitudinal $z$-component.
However, we keep in mind the general case of {\it nonsymmetric potentials} (especially 
with respect to the inversion $x \ra -x$ at the location of dipoles, because of which 
the polarization may have a transverse component.    

Hamiltonian (\ref{1}) is the standard, widely used, Hamiltonian for describing 
macroscopic ferroelectric samples. For finite samples, in general, one should take into 
account the depolarizing field caused by the charges on the surfaces of the sample 
\cite{Watanabe_61}. However, there are ways \cite{Rabe_62} of compensating 
surface charges, thus reducing or removing depolarizing fields. 

Also, considering an external electric field $E_0$, applied in the $z$-direction, 
leads to the appearance of the depolarizing field proportional to $-P_z$, which 
is of the order of $\rho d_0 S$, where $\rho$ is the sample density. The energy, 
corresponding to the depolarizing field, is $d_o P_z$, which gives $\rho d_0^2 S$. 
The latter expression equals the dephasing rate or transverse attenuation denoted 
by $\gamma_2 = \rho d_0^2 S$. The magnitude of the energy, corresponding to the 
external field, is $|d_0 E_0|$, which defines the dipole rotation frequency denoted 
as $\omega_0 = |d_0 E_0|$. For what follows, we will need a sufficiently strong 
external field, such that $\omega_0$ be much larger than $\gamma_2$. This is necessary 
for realizing coherent motion of dipoles, so that the reversal time be much shorter 
than the dephasing time. Under the condition $\omega_0 >> \gamma_2$, corrections, 
related to the depolarizing field, can be omitted in the Hamiltonian. At the same 
time, the attenuation rate $\gamma_2$ will be taken into account in the equations of 
motion.

The total electric field consists of two terms,
\be
\label{3}
\bE_{tot} = E\bfe_x + E_0 \bfe_z \;  ,
\ee
where the first term is a field of the resonator cavity, in which the sample is inserted, 
and the second term is an external constant electric field. The resonator cavity is chosen 
such that it supports the TM$_{010}$ fundamental mode, whose electric field is directed 
along the cavity axis that is taken to be the $x$-axis. The resonator cavity electric 
field is a feedback field generated by the moving polarization 
\be
\label{4}
 \bP = \frac{d_0}{V} \sum_j \; \lgl \; \bS_j \; \rgl \;  ,
\ee
with $V$ being the sample volume. 

The equation for the feedback field can be derived in the standard way \cite{Mandel_26}.
From the Maxwell equations inside the cavity with an inserted ferroelectric, it is 
straightforward to get the equation for the electric field
\be
\label{E1}
 \nabla^2\bE \; - \; \frac{1}{c^2}\; \frac{\prt^2\bE}{\prt t^2} \; - \;
 \frac{4\pi\sgm}{c^2}\; \frac{\prt\bE}{\prt t}  = 
\frac{4\pi}{c^2}\; \frac{\prt^2\bP}{\prt t^2}
\ee
generated by the ferroelectric polarization (\ref{4}), where $\sigma$ is conductivity and
$c$ is light velocity. It is possible to look for the solution to this equation in the form
\be
\label{E2}
 \bE(\br,t) = {\bf e}(\br) E(t) \;  ,
\ee
where ${\bf e}({\bf r})$ is a cavity mode defined by the Helmholtz equation and normalized
to the cavity volume $V_c$, so that
\be
\label{E3}
\frac{1}{V_c} \int |\; {\bf e}(\br) \; |^2 \; d\br = 1 \;   .
\ee
We are looking for the TM$_{010}$ fundamental mode, which, by definition, is the mode 
directed along the cavity axis that here is the axis $x$, which implies the conditions
\be
\label{E4}
e_y(\br) = 0 \; , \qquad e_z(\br) = 0 \;   .
\ee
The mode $x$-component is nonzero inside the cavity, while satisfies the boundary condition
\be
\label{E5}
e_x(\br)|_{r=R} = 0 
\ee
on the cavity cylindrical surface of radius $R$. The expression for the TM$_{010}$ 
fundamental mode is known \cite{Mandel_26} to be presented through the Bessel function of 
the first kind,
$$
e_x({\bf r}) = C_0 J_0 \left( \frac{\om}{c}\; r\right) \; , \qquad 
C_0 = J_1^{-1} \left( \frac{\om}{c}\; R\right) \; , 
$$
with $C_0$ being the normalization constant. The boundary condition $e_x(R) = 0$ corresponds 
to the first zero of the Bessel function $J_0(\omega R/ c) = 0$, which defines the cavity 
natural frequency 
\be
\label{E6}
\om = 2.4048\; \frac{c}{R} \;   .
\ee
The normalization constant becomes
$$
C_0 = \frac{1}{J_1(2.4048)} = \frac{1}{0.51915} = 1.9262 \;  .
$$
Thus the TM$_{010}$ fundamental mode reads as
\be
\label{E7}
  {\bf e}(\br) = C_0 J_0 \left( \frac{\om}{c}\; r\right) {\bf e}_x = 
e_x(\br) {\bf e}_x \; ,
\ee
where ${\bf e}_x$ is a unit vector along the cavity axis $x$. Then we substitute 
expression (\ref{E2}), with mode (\ref{E7}), into Eq. (\ref{E1}), multiply the latter 
by form (\ref{E7}), take into account that 
$$
{\bf P} \cdot {\bf e}({\bf r}) = P_x e_x(\br) \; ,
$$
and integrate over the cavity volume. This leads to the equation 
\be
\label{5}
 \frac{d^2E}{dt^2} + 2\gm \; \frac{dE}{dt} + \om^2 E = 
- 4\pi \eta_f \; \frac{d^2P_x}{dt^2} \; ,
\ee
in which $\gamma = 2 \pi \sigma$ is a cavity attenuation, $\omega$ is the cavity natural 
frequency (\ref{E6}), and 
$$
\eta_f \equiv \frac{1}{V_c} \int e_x(\br) \; d\br = 0.83167
$$
is the filling factor corresponding to the TM$_{010}$ fundamental mode.      

The evolution equations, following from the Heisenberg equations of motion are
$$
\frac{dS_i^x}{dt} = \left( \sum_j J_{ij} S_j^z + d_0 E_0 \right) S_i^y \; ,
$$
$$
 \frac{dS_i^y}{dt} = - \left( \sum_j J_{ij} S_j^z + d_0 E_0 \right) S_i^x +
(\Om + d_0 E ) S_i^z \; ,
$$
\be
\label{6}
\frac{dS_i^z}{dt} = - (\Om + d_0 E ) S_i^y \;  .
\ee
The observable quantities are given by the statistical averages
\be
\label{7}
 s_\al \equiv \frac{1}{NS} \sum_j \; \lgl \; S_j^\al \; \rgl \qquad
( \al = x,y,z ) \;  ,
\ee
where $N$ is the number of lattice sites and $S = 1/2$ is the spin value. 

The attenuation can be taken into account by employing the method of local fields
\cite{Wangsness_27,Yukalov_28}, where the attenuation is caused by particle interactions
acting in the local field formed by other particles, so that the dynamic variables are
forced to relax to their local equilibrium values. In the present case, the latter are
\be
\label{8}
 \zeta _\al \equiv   \frac{1}{NS} \sum_j \; \lgl \; S_j^\al \; \rgl_{loc} \; ,
\ee
where $S = 1/2$ and the local equilibrium averages are expressed through variables 
(\ref{7}) taken at the given moment of time. Since dipolar interactions are of long range, 
the mean-field approximation is applicable. In this way, the quantity 
$\lgl \; S_j^\al \; \rgl_{loc}$ is defined as the average
\be
\label{loc1}
\langle \; S_j^\alpha \; \rangle_{loc} = {\rm Tr}\; \hat\rho_{loc}\; S_j^\al \;  ,
\ee
with the local equilibrium statistical operator
$$
{\hat \rho}_{loc} =  \frac{\exp(-\hat H_{loc}/T)}{{\rm Tr}\exp(-\hat H_{loc}T)}   
$$
for the ensemble of spins with the Hamiltonian
$$
{\hat H}_{loc} = -\Om \sum_j S_j^x \; - \; 
s_z \sum_{i\neq j} J_{ij}S_j^z \; - \; d_0 E_0 \sum_j S_j^z \;   .
$$
Accomplishing explicit calculations for the local average (\ref{loc1}), we keep in
mind low temperatures, such that $T \ll J$, where
\be
\label{loc2}
J \equiv \frac{1}{N} \sum_{i\neq j} J_{ij} \; .
\ee    
Thus we obtain the low-temperature local equilibrium values
$$
\zeta_x = \frac{\Om}{[\Om^2+\om_0^2(1-As_z)^2]^{1/2}} \; , \qquad \zeta_y = 0 \; ,
$$
\be
\label{9}
\zeta^2_x + \zeta^2_z = 1\; , \qquad
\zeta_z = -\; \frac{\om_0(1-As_z)}{[\Om^2+\om_0^2(1-As_z)^2]^{1/2}} \;  ,
\ee
where we introduce the notation
\be
\label{10}
 A \equiv \frac{JS}{\om_0}   
\ee
and define the frequency
\be
\label{11}
 \om_0 \equiv - d_0 E_0 > 0 \;  .
\ee
The positive value of $\omega_0$ implies that the external electric field is directed 
downwards.

Thus we come to the mean-field evolution equations for the pseudospin variables
$$
\frac{ds_x}{dt} = -\om_0 ( 1 - As_z ) s_y - \gm_2 ( s_x - \zeta_x) \; ,
$$
$$
\frac{ds_y}{dt} = \om_0 ( 1 - As_z ) s_x + ( \Om + \gm_2 h) s_z - \gm_2 s_y \; ,
$$
\be
\label{12}
\frac{ds_z}{dt} = - ( \Om + \gm_2 h) s_y - \gm_1 ( s_z - \zeta_z)\;   .
\ee
Here $\gamma_1$ is the longitudinal relaxation rate due to spin-phonon interactions
(see Blinc \cite{Blinc_24}), while $\gamma_2 = \rho d_0^2 S$ is the transverse attenuation 
caused by dipolar interactions. Usually, $\gamma_1 \ll \gamma_2$. 

By introducing the dimensionless feedback field
\be
\label{13}
 h \equiv \frac{d_0E}{\gm_2} \qquad ( \gm_2 = \rho d_0^2 S ) \; ,
\ee
where $\rho = N/V$, and taking account of the expression $P_x=\rho d_0 S s_x$, we get 
the feedback-field equation
\be
\label{14}
 \frac{d^2h}{dt^2} + 2\gm\; \frac{dh}{dt} + \om^2 h = 
- 4\;\frac{\gm_f}{\gm_2} \; \frac{ds^2_x}{dt^2}  \; .
\ee
Here 
\be
\label{15}
\gm_f \equiv \pi \eta_f \rho d_0^2 S = \pi \eta_f \gm_2 = 2.6 \gm_2
\ee
is the coupling rate characterizing the interaction between the ferroelectric sample and 
resonator. 

Equations (\ref{12}) and (\ref{14}) define the dynamics of the variables $s_\alpha$ and the
feedback field $h$. The most interesting is the behavior of the dimensionless polarization
$s_z = s_z(t)$ as a function of time for different parameters, under the given initial 
polarization $s_z(0) = s_0$.

\section{Numerical solution}

The polarization switching is realized in the following way. Suppose the ferroelectric 
sample is initially polarized along the axis $z$. The sample is placed inside a resonator 
cavity supporting the TM$_{010}$ fundamental mode and is subject to an external electric 
field directed opposite to the initial polarization. The polarization dynamics is governed 
by Eqs. (\ref{12}) and (\ref{14}). In order to precisely describe this dynamics, we have 
to fix realistic parameters typical for ferroelectrics \cite{Uchino_1,Dawber_2,Blinc_24}.

First, let us notice that a ferroelectric characterized by Hamiltonian (\ref{1}), without 
the last term containing electric fields, acquires spontaneous polarization below the 
critical temperature
\be
\label{16}
T_c = \Om \left( \ln \; \frac{J+2\Om}{J-2\Om} \right)^{-1} \;  .
\ee
This temperature is positive for the tunneling frequency
\be
\label{17}
\Om < \frac{J}{2} \qquad ( T_c > 0 ) \;  .
\ee

The interaction strength, due to dipolar forces, $J \approx \rho d_0^2$. The electric 
dipole is $d_0=e_0l_0$, with $l_0\sim 10^{-8}$ cm and the electric charge about a proton 
charge $e_0=1.602\times 10^{-19}$ C. Keeping in mind that one Coulomb $1C=2.998\times 10^9$
g$^{1/2}$ cm$^{3/2}/$s, we find $d_0 \sim 10^{-27}$ C cm $\sim 1D$, where one Debye is 
$1D=3.336\times 10^{-28}$ C cm. For the density $\rho\sim 10^{22}$ cm$^{-3}$, we obtain 
$\rho d_0^2\sim 10^{-14}$ erg, that is $\rho d_0^2\sim 10^{13}$ s$^{-1}$. Really, for 
typical ferroelectrics $J\sim 10^2$ K, that is, $J\sim 10^{13}$ s$^{-1}$. The dipolar 
forces induce the transverse attenuation $\gm_2=\rho d_0^2 S$. Thus we have 
$$
 JS \approx \gm_2 = \rho d_0^2 S \sim 10^{13} {\rm s}^{-1} \;  .
$$
The longitudinal attenuation, caused by the interaction of pseudospins with phonons, is 
much smaller than the transverse attenuation, $\gamma_1 \ll \gamma_2$. 
   
The cavity is called resonant, since its natural frequency $\omega$ has to be tuned close 
to the dipole rotation frequency $\omega_0$, satisfying the quasi-resonance condition
\be
\label{18}
 \left| \; \frac{\Dlt}{\om} \; \right| \ll 1 \qquad ( \Dlt \equiv \om - \om_0 ) \;  .
\ee  
While the attenuations are to be smaller than $\omega_0$, so that
\be
\label{19}
 \frac{\gm_2}{\om_0} \ll 1 \; , \qquad \frac{\gm}{\om} \ll 1 \;  .
\ee
Therefore, for parameter (\ref{10}), we get
\be
\label{20}
 A \equiv \frac{JS}{\om_0} = \frac{\gm_2}{\om_0} \ll 1 \;  .
\ee

Since $JS=\gm_2 \sim 10^{13}$ s$^{-1}$, to satisfy condition (\ref{20}), we need that 
$\om_0$ be at least about $10^{14}$ s$^{-1}$ to $10^{15}$ s$^{-1}$, which is in the 
near infrared or visible light range. This gives the wave vector $k_0\equiv\om_0/c$ 
of the order $3\times 10^3$ cm$^{-1}$ to $3\times 10^4$ cm$^{-1}$ and the wavelength 
$\lbd\sim 10^{-4}$ cm to $10^{-3}$ cm. Resonator cavities in the range of visible light 
are widespread, and also there exist various cavities operating in the infrared region 
\cite{Garin_29,Lecaplain_30,Osman_31,Radosavljevic_32,Yao_33,Xiao_34}. 

We solve numerically the system of Eqs. (\ref{12}) and (\ref{14}), concentrating our 
attention on the behavior of the polarization $s_z=s(t)$ as a function of time, for 
different parameters in the admissible range. In the figures, time is measured in units 
of $1/\gm_2$ and the frequency parameters are measured in units of $\gm_2$. As initial 
conditions, we need to fix the values $s_x(0) = \sqrt{1 - s_0^2}$, $s_y(0)$, 
$s_z(0)\equiv s(0)\equiv s_0$, $h(0)$, and the time derivative $\dot{h}(0)$. The initial 
polarization is positive, $s_0>0$. If some of other initial conditions, except $s_0$, 
are not zero, the reversal begins immediately at $t=0$. When all of them (except $s_0$) 
are zero, there is a time delay. In the figures, we show the results for the initial 
conditions $s_y(0)=0$, $h(0)=0$, and $\dot{h}(0)=0$, while $s_0$ and, respectively, 
$s_x(0)=\sqrt{1-s_0^2}$ can be varied. The resonance is assumed, when $\om=\om_0$. 

Figure 1 demonstrates the polarization reversal at different values of the resonator 
attenuation. For small $\gamma$, the polarization oscillates after the switching. Hence, 
in order to achieve a steady state after the polarization reversal, it is necessary to 
take larger $\gm$. For $\gm=10$, the after-switching oscillations are suppressed. Thus,
to avoid oscillations, it is preferable that the resonator ringing time $\tau\equiv 1/\gm$
be shorter than the dephasing time $T_2\equiv 1/\gm_2$.  
  
Figure 2 shows the dependence of the polarization switching on the tunneling frequency. 
The larger $\Omega$, the shorter the delay time. As is clear from the evolution equations,
this is because the tunneling triggers the polarization motion. 

In Fig. 3, the role of the frequency $\omega$ is illustrated. The larger $\omega$, the 
shorter the delay time and better the polarization inversion. This happens because the 
larger frequency makes stronger the coupling between the resonator cavity and the 
ferroelectric sample.

Figure 4 shows the similar dependence of the polarization switching on the frequency 
$\omega$, as in the previous figure, but for the initial polarization $s_0 = 0.5$, when 
$s_x(0)=\sqrt{1-s_0^2}$ is not zero. As is mentioned above, a nonzero $s_x(0)$ triggers 
the start of the polarization motion, so that there is no delay time, and the switching 
begins from $t = 0$.

In Fig. 5, the polarization switching for different initial polarizations is compared: 
$s_0 = 0.5$ (solid line) and $s_0 = 1$ (dashed-dotted line). For $s_0 = 0.5$, the initial
transverse component $s_x(0) = \sqrt{1 - s_0^2}$ is not zero, because of which the process 
of switching starts from the very beginning, practically at $t = 0$, without delay.    

Figure 6 demonstrates that the transverse polarization component $s_x$ oscillates around 
zero. The oscillation is faster for larger $\omega$. The maximal oscillation amplitude
corresponds to the moment of the polarization switching. 

Since the transverse component $s_x$ generates, by means of relation (\ref{14}), the 
dimensionless cavity field $h$, hence the dimensional electric field inside the cavity $E$, 
the temporal behavior of $h$, as is seen from Fig. 7, is similar to that of the component 
$s_x$.

\section{Analytical solution}

Although the numerical solution of the previous section gives us an accurate description 
of the process of polarization switching, nevertheless it is desirable to have analytic
solutions that would provide, at least approximately, explicit formulas allowing for the 
better understanding of the related physics and for straightforward estimates of 
characteristic quantities. 

It is convenient to pass to the variables
$$
 u \equiv s_x - i s_y \; ,
$$
\be
\label{21}
w \equiv |\; u\; |^2 =  s_x^2 + s_y^2 \; , 
\qquad  s \equiv s_z \;  .
\ee
In terms of these variables, Eqs. (\ref{12}) transform into the equation for the 
transverse component
\be
\label{22}
 \frac{du}{dt}= 
- i\om_0 ( 1 - As) u - \gm_2 u - 
i\gm_2\left( h + \frac{\Om}{\gm_2} \right) s + \gm_2 \zeta_x \;  ,
\ee
for the coherence intensity
\be
\label{23}
\frac{dw}{dt}= 
- 2\gm_2 w  - i\gm_2\left( h + \frac{\Om}{\gm_2} \right)\left( u^* - u \right)  s + 
\gm_2 \zeta_x \left( u^* + u \right) \;   ,
\ee
and for the polarization,
\be
\label{24}
\frac{ds}{dt}= 
\frac{i}{2} \; \gm_2\left( h + \frac{\Om}{\gm_2} \right)\left( u^* - u \right)
- \gm_1  ( s - \zeta_z ) \;   .
\ee

Keeping in mind the case of resonance, defined by Eq. (\ref{18}), and the existence of the
small parameters described in Eqs. (\ref{17}), (\ref{19}), and (\ref{20}), we notice that 
the variables $u$ and $h$ can be classified as fast, while the variables $w$ and $s$, as 
slow. In that case, for solving the given system of equations, it is admissible to resort 
to the averaging techniques \cite{Bogolubov_35,Burd_36}. Below, we follow the variant of 
the method described in detail in Refs. \cite{Yukalov_37,Yukalov_38}. First, we solve the 
equations for the fast variables, keeping there the slow variables as quasi-integrals of 
motion. Such a solution is straightforward, although cumbersome, since the equations for 
the fast variables become linear with respect to the latter, when the slow variables are 
kept fixed. Then the found solutions for the fast variables are substituted into the 
equations for the slow variables, with the averaging of the slow-variable equations over 
time. This yields the equations for the guiding centers, which can be analyzed. All this 
machinery has been thoroughly described in Refs. \cite{Yukalov_37,Yukalov_38} and its use 
has been demonstrated for studying the dynamics of magnetic systems 
\cite{Yukalov_39,Yukalov_40,Yukalov_41}.

To present the solutions for the fast variables, under fixed slow variables, in a compact
form, we take account of the small parameters and introduce several notations. We define 
the coupling parameter 
\be
\label{25}
g \equiv \frac{\gm_f\om_0}{\gm\gm_2} =  2.6 \;\frac{\om_0}{\gm}
\ee
characterizing the strength of the coupling between the ferroelectric sample and resonator,
the coupling function
\be
\label{26}
\al \equiv g ( 1 - As ) \left( 1 - e^{-\gm t} \right)
\ee
describing the dynamics of the ferroelectric-resonator interaction, and the effective 
frequency
\be
\label{27}
\om_{eff} \equiv \om_0 ( 1 - A s) - i \gm_2 ( 1 - \al s ) \; .
\ee
Thus we obtain the transverse component
\be
\label{28}
u = \left( u_0 + \frac{\Om s + i\gm_2\zeta_x}{\om_{eff} } \right) \exp(- i\om_{eff} t )
- \; \frac{\Om s + i\gm_2\zeta_x}{\om_{eff}}
\ee
and the feedback field
\be
\label{29}
  h = - i\al \left( u^* - u \right) \; .
\ee
Note that from Eq. (\ref{9}), we have
$$
\zeta_x \cong \frac{\Om}{\om_0} \; , \qquad \zeta_z \cong - 1 \;  .
$$
    
Substituting the fast variables into the equations for the slow variables, and averaging 
the resulting equations over time gives the guiding-center equations for the coherence
intensity,
\be
\label{30}
\frac{dw}{dt} = - 2 \gm_2 ( 1 - \al s) w + 2 \gm_3 ( 1 - \al - \al s ) s^2 \; ,
\ee
and for the polarization
\be
\label{31}
\frac{ds}{dt} = - \gm_2 \al w - \gm_3 ( 1 + s - 2\al s) - \gm_1 ( s - \zeta_z) \; ,
\ee
where
\be
\label{32}
 \gm_3 \equiv \gm_2 \; \frac{\Om^2}{\om_0^2} \; .
\ee

The parameter $\gamma_3$ is very small. However, it cannot be neglected, since it plays 
an important role in triggering the polarization motion at the initial stage. At the very 
beginning of the process, when $t \ra 0$, so that
\be
\label{33}
 \gm t \ll 1 \; , \qquad   \gm_1 t \ll 1 \; , \qquad  \gm _2t \ll 1 \; , \qquad
\gm _3t \ll 1 \; ,
\ee
the coupling function (\ref{26}) is close to zero. Then, keeping in mind that usually 
$\gamma_1 \ll \gamma_2$, the equations of motion become
$$
\frac{dw}{dt} = - 2 \gm_2 w + 2\gm_3 s^2 \; ,
$$
\be
\label{34}
\frac{ds}{dt} = -  \gm_3 ( 1 + s ) \qquad ( t \ra 0 ) \;   .
\ee
Their solutions are
$$
w \simeq \left( w_0 \; - \; \frac{\gm_3}{\gm_2} \; s_0^2 \right) e^{-2\gm_2 t} +
 \frac{\gm_3}{\gm_2} \; s_0^2 \; , 
$$
$$
s \simeq  ( 1 + s_0 ) e^{-\gm_3 t} - 1 \; ,
$$
which, in view of inequalities (\ref{33}), can be simplified to
$$
w \simeq w_0 + 2 \left( \gm_3 s_0^2 - \gm_2 w_0 \right) t \; ,
$$
\be
\label{35}
 s \simeq s_0 - \gm_3 (  1 + s_0 ) t \; ,
\ee
where $w_0 \equiv w(0) = 1- s_0^2$ and $s_0 \equiv s(0)$. These are the solutions at the 
initial stage, when the motion of individual polarizations is not mutually synchronized.  

The coupling function (\ref{26}) grows with time, implying the increase of the magnitude
of the resonator feedback field, which collectivizes the individual polarizations, forcing
them to move coherently. The influence of the feedback field becomes crucial after the
{\it coherence time} $t_{coh}$, when the coupling function grows so that
\be
\label{36}
 \al s = 1 \qquad ( t = t_{coh} ) \; .
\ee
This defines the coherence time
\be
\label{37}
t_{coh} = \tau \ln \; \frac{gs_0(1 - As_0 )}{gs_0(1 - As_0 )-1} \qquad 
\left( \tau \equiv \frac{1}{\gm} \right)
\ee
with $\tau$ being the resonator ringing time. If the ferroelectric-resonator coupling 
is strong, such that $g s_0 \gg 1$, then the coherence time is
\be
\label{38}
 t_{coh} \simeq  \frac{\tau}{gs_0(1 - As_0 )} \;  .
\ee

At the coherence time, the solutions (\ref{35}), that is 
\be
\label{39}
 w_{coh} = w(t_{coh} ) \; ,   \qquad    s_{coh} = s(t_{coh} ) \;  ,
\ee      
can be written as
\be
\label{40}
w_{coh} \simeq w_0 +2\gm_3 s_0^2 t_{coh} \; , \qquad   s_{coh} \simeq s_0 \; ,
\ee
where we assume that the coupling parameter $g$ is sufficiently large, so that 
inequalities (\ref{33}) are yet valid at $t_{coh}$. 

After $t_{coh}$, the coupling function (\ref{26}) quickly grows reaching the value 
$g(1-As)\simeq g$. At this stage, the parameters $\gm_1$ and $\gm_3$, that are much 
smaller than $\gm_2$, and especially than $g\gm_2$, can be neglected. Then Eqs. (\ref{30}) 
and (\ref{31}) become
$$
\frac{dw}{dt}= - 2\gm_2 ( 1 - gs ) w_0 \; ,
$$
\be
\label{41}
\frac{ds}{dt}= - \gm_2  g w  \qquad ( t > t_{coh} ) \;   .
\ee
The latter equations enjoy the exact solutions for the coherence intensity
\be
\label{42}
w = \left( \frac{\gm_s}{g\gm_2} \right)^2 
{\rm sech}^2\left( \frac{t-t_0}{\tau_s}\right)
\ee
and polarization
\be
\label{43}
s = -\;  \frac{\gm_s}{g\gm_2}\;
{\rm tanh} \left( \frac{t-t_0}{\tau_s}\right) + \frac{1}{g} \;  .
\ee
The quantities $\gamma_s$ and $t_0$ are the integration constants that are defined by 
sewing expressions (\ref{42}) and (\ref{43}) with Eqs. (\ref{40}). This gives the 
{\it switching time} $\tau_s \equiv 1/\gamma_s$, in which 
\be
\label{44}
 \gm_s^2 = \gm_g^2 + (g\gm_2)^2 w_{coh} \; , \qquad 
\gm_g \equiv \gm_2 ( gs_0 -1 ) \;  ,
\ee
and the {\it delay time}
\be
\label{45}
 t_0 = t_{coh} + 
\frac{\tau_s}{2}\; \ln \left( \frac{\gm_s+\gm_g}{\gm_s-\gm_g}\right) \;  .
\ee
The delay time shows the time, when the switching starts, while the switching time is the
time during which the polarization reversal occurs. The delay time can also be written as
\be
\label{46}
 t_0 = t_{coh} + 
\tau_s \ln \left( \frac{\gm_s+\gm_g}{g\gm_2\sqrt{ w_{coh} } }\right) \;  ,
\ee
which demonstrates its explicit dependence on $w_{coh}$.

\section{Characteristic quantities} 

The derived analytic expressions provide a transparent illustration for the role of 
different system parameters and their combinations. To study more carefully these
dependencies, let us keep in mind the case of strong ferroelectric-resonator coupling,
when $g s_0\gg 1$. Then the coherence time (\ref{38}) takes the form
\be
\label{47}
 t_{coh} \simeq \frac{0.385}{\om_0 s_0} \; ,
\ee
which shows that the larger the frequency $\om_0$, the shorter this time. 

Quantities (\ref{44}) read as
\be
\label{48}
\gm_s \simeq g\gm_2 \left( 1  + 0.385\; \frac{\gm_2\Om^2}{\om_0^3}\; s_0 \right) \; ,
\qquad \gm_g \simeq \gm_2 g s_0 \;  .
\ee
The switching time $\tau_s \equiv 1/\gamma_s$ becomes
\be
\label{49}
 \tau_s \simeq 0.385\; \frac{\gm}{\om_0\gm_2} \;  ,
\ee
so that larger $\omega_0$ makes the process of switching faster. The delay time (\ref{46}) 
acquires the form
\be
\label{50}
 t_0 \simeq  t_{coh} + \tau_s \ln \left( \frac{1+s_0}{\sqrt{w_{coh}} } \right) \; ,
\ee
which depends on the value $w_{coh}$. The latter is connected with the initial polarization
$s_0$. 
 
When the system at the initial moment of time is well polarized, with $s_0=1$ and $w_0= 0$,
then 
\be
\label{51}
 w_{coh} = 0.77 \; \frac{\gm_2\Om^2}{\om_0^3} \qquad ( s_0 = 1 ) \;  .
\ee
And the delay time turns into
\be
\label{52}
 t_0 \simeq \frac{0.385}{\om_0} \left[ 1 + 
\frac{\gm}{2\gm_2}\;\ln\left( \frac{5.2\om_0^3}{\gm_2\Om^2}\right) \right] \qquad
( s_0 = 1 ) \;  .
\ee
While when $s_0=0.5$ and $w_0=0.75$, then the delay time is
\be
\label{53}
 t_0 \simeq \frac{0.77}{\om_0} \left( 1 + 0.65\; \frac{\gm}{\gm_2} \right) \qquad 
( s_0 = 0.5 ) \; .
\ee

To get concrete estimates, let us take the typical values of parameters as have been used
when numerically solving the evolution equations: $\om_0\sim 100\gm_2$, $\Om\sim 0.1\gm_2$, 
$\gm\sim 10 \gm_2$, and $s_0\sim 1$. Then $g\sim 10$, $\gm_s\sim \gm_g \sim 10\gm_2$, and 
$\gamma_3 \sim 10^{-6} \gamma_2$. For the coherence time, we have $t_{coh} \sim 10^{-15}$ s 
and for the switching time we get $\tau_s \sim 10^{-14}$ s. Diminishing $s_0$ decreases the 
delay time. Thus, if $s_0=1$ and $w_0=0$, then $w_{coh}\sim 10^{-8}$ and $t_0\sim 10^{-13}$ s. 
But if $s_0 \approx 0.5$, so that $w_0 \sim 1$, then $t_0 \sim 10^{-14}$ s. These estimates 
are in good agreement with numerical calculations. 

An important question is how the switching time is limited in realistic materials. In 
particular, what is the relation between the switching time $\tau_s$ and the cavity ringing
time (delay time $\tau \equiv 1/ \gamma$). From the above estimates, we find the ratio
\be
\label{R1}
\frac{\tau_s}{\tau} = 0.385\; \frac{\gm^2}{\om_0\gm_2} \;  .
\ee 
It looks that by varying the system parameters, it is possible to make this ratio rather 
small. However, there are limitations for the variation of the parameters. Thus, for a good 
quality cavity one has $\gamma \ll \omega_0$. But $\gamma$ cannot be arbitrarily small, since 
for $\gamma < \gamma_2$, there appear oscillations in the polarization. In order to realize 
a stable switching without oscillations, it is necessary to take $\gamma \gg \gamma_2$. 
Therefore ratio (\ref{R1}) lies in the interval
\be
\label{R2}
\frac{\gm_2}{\om_0} \ll  \frac{\tau_s}{\tau} \ll \frac{\om_0}{\gm_2} \qquad
(\gm_2 \ll \gm \ll \om_0 ) \; .
\ee
For the infrared region, where $\omega_0/ \gamma_2 \sim 10$, we have
$$
0.1 \ll \frac{\tau_s}{\tau} \ll 10 \qquad 
\left( \frac{\om_0}{\gm_2} \sim 10 \right) \; ,
$$ 
which actually means that the switching time $\tau_s$ is of order of the ringing time $\tau$. 
In the visible light region, when $\omega_0/ \gamma_2 \sim 100$, we find
$$
0.01 \ll  \frac{\tau_s}{\tau} \ll 100 \qquad 
\left( \frac{\om_0}{\gm_2} \sim 100 \right) \; .
$$
This tells us that again the switching time is correlated with the ringing time. For the 
visible light, it looks admissible to reach the shortest switching time of order
$\tau_s \sim 10^{-15}$ s.

\section{Coherent radiation}
 
The motion of electric dipoles has to produce electromagnetic radiation. If this motion is
coherent, the produced radiation should also be coherent. Since the sample is inside a 
resonator cavity, the radiation can propagate only along the cavity axis, that is, along 
the axis $x$. The radiation intensity in the direction of ${\bf n} = {\bf e}_x$ consists
of two terms describing incoherent and coherent radiation,
\be
\label{54}
  I(\bn,t) = I_{inc}(\bn,t) + I_{coh}(\bn,t) \; .
\ee
Radiation, produced by moving dipoles can be described in the following way
\cite{Mandel_26,Yukalov_38,Gross_42}. The incoherent radiation intensity reads as
\be
\label{55}
 I_{inc}(\bn,t) = \frac{3}{16\pi}\; N \om_0 \gm_0 [ 1 + s(t) ]  
\ee
and the coherent radiation intensity is
\be
\label{56}
 I_{coh}(\bn,t) = \frac{3}{32\pi}\; N^2 \om_0 \gm_0 w(t) F(k_0\bn) \;  ,
\ee
with the shape factor
\be
\label{57}
 F(k_0\bn) = \frac{4}{k_0^2L^2} \; \sin^2\left( \frac{k_0L}{2}\right) \;  ,
\ee
where $L$ is the cavity length and $\gamma_0$ is the natural width
\be
\label{58}
 \gm_0 = \frac{2}{3} \; |\; \bd_0\; |^2 k_0^3 \qquad 
\left( k_0 = \frac{\om_0}{c} \right) \;  .
\ee

For the frequency $\om_0 \sim 10^{15}$ s$^{-1}$, we have the wavelength 
$\lbd\sim 10^{-4}$ cm and the natural width $\gamma_0 \sim 10^4 {\rm s}^{-1}$. Then the 
radiation intensities at the maximum, are
$$
I_{inc}(\bn,t) \sim N 10^{-15} W \; , 
$$
$$
I_{coh}(\bn,t) \sim N^2 F(k_0\bn) 10^{-16} W \;   .
$$
The number of dipoles that could radiate coherently can be estimated as $N\sim \rho\lbd^3$.
If we consider a small sample, with the length $L$ smaller than the radiation wavelength 
$\lbd$, then the shape factor (\ref{57}) is of order one. In that case, for the density 
$\rho\sim 10^{22}$ cm$^{-3}$, we get $N\sim 10^{10}$. And for the radiation intensities, 
we find
$$
 I_{inc}(\bn,t) \sim 10^{-5} W \; , \qquad I_{coh}(\bn,t) \sim 10^4 W \;  .
$$

When the frequency is $\omega_0 \sim 10^{14} {\rm s}^{-1}$, then $\lambda \sim 10^{-3}$ cm
and $\gamma_0 \sim 10 {\rm s}^{-1}$. The radiation intensities are
$$
I_{inc}(\bn,t) \sim N 10^{-20} W \; , 
$$
$$
I_{coh}(\bn,t) \sim N^2 F(k_0\bn) 10^{-21} W \;   .
$$
Considering again coherently radiating dipoles, with the number $N \sim \rho \lambda^3$, we 
have $N \sim 10^{13}$. This gives for the radiation intensities
$$
I_{inc}(\bn,t) \sim 10^{-7} W \; , \qquad I_{coh}(\bn,t) \sim  10^5 W \;   .
$$
Such a level of radiation can be easily measured. The prevailing coherent component of 
radiation shows that this radiation is of the type of superradiance.

\section{Conclusion}

A method of ultrafast polarization switching in ferroelectrics is suggested. The main idea
is to place a ferroelectric sample into a resonator cavity. In the presence of a constant 
electric field, directed opposite to the ferroelectric polarization, the sample is in a 
nonequilibrium state. As soon as the polarization starts moving, it produces an electric 
field in the cavity. This field acts back on the sample forcing the polarization to move 
faster. Thus the ferroelectric itself generates a feedback field accelerating the 
polarization motion, so that there is no necessity of applying external alternating fields, 
as one usually does for realizing polarization switching. It turns out that the 
self-organized feedback field is essentially more effective for the polarization reversal 
than an externally imposed field. 

The system of equations, describing the ferroelectric polarization and feedback field is
solved numerically and also analytically by means of averaging techniques. This makes it 
possible to give a detailed description of the whole procedure, to study the role of the 
system parameters, and to estimate the characteristic quantities involved in the process. 
The polarization switching can be realized extremely fast: for the parameters of typical 
ferroelectrics, the switching time can reach femtoseconds. This ultrafast polarization 
reversal generates a coherent electromagnetic pulse.        

We stress that the main goal of the paper is to attract attention to the possibility of 
accelerating the polarization switching in ferroelectrics by using the self-acceleration 
effect caused by the action of a resonator-cavity feedback field. This is the idea of 
principle that, to our knowledge, has not been considered for ferroelectrics before. 
It goes without saying that the method is not necessarily applicable to any particular 
material. Thus the method seems to be not applicable for the order-disorder 
ferroelectrics with spatially symmetric double wells, where there is only the longitudinal 
polarization. However the considered model assumes the general case of asymmetric 
potentials for which the method is applicable. We also hope that, since there are various 
types of ferroelectric systems \cite{Blinc_24,Lines_4,Strukov_5,Whatmore_6}, there can 
exist other materials for which the suggested idea could work.  

In order that the suggested method could be realized, the existence of two spin
components of polarization, longitudinal and transverse, is required. If one keeps 
in mind an order-disorder ferroelectric with lattice-site double wells that are 
ideally symmetric with respect to spatial inversion (especially with respect to the 
inversion $x \ra -x$), then there is only a longitudinal component, and the sample 
polarization is expressed through the $z$-component of the spin operator. But in the 
general case of an asymmetric double well, the sample polarization contains a term with 
the $x$-component of spin. The asymmetry can be induced by stress or by incorporating 
into the sample admixtures or vacancies. Thus, the inclusion of vacancies in order-disorder 
ferroelectrics is attributed to the breaking of spatial inversion symmetry along different 
directions \cite{Yoo_8}. The symmetry in order-disorder ferroelectrics can be distorted 
by the action of a transverse electric field \cite{Fugiel_9}.

To illustrate an idea, one has to consider some model. We considered an Ising-type model 
in a transverse field. This kind of models, employing the spin representation, is widely 
used for order-disorder ferroelectrics \cite{Blinc_24,Lines_4,Strukov_5,Whatmore_6} and 
relaxor ferroelectrics \cite{Bokov_25,Hong_7}.

In addition to different ferroelectrics 
\cite{Blinc_24,Lines_4,Strukov_5,Whatmore_6,Hong_7} and multiferroics 
\cite{Wang_11,Xiang_12}, ferroelectric-type spin models are widely used for 
describing the systems of polar molecules, Rydberg atoms, Rydberg-dressed atoms, 
dipolar ions, vacancy centers in solids, and quantum dots 
\cite{Baranov_13,Krems_14,Dulieu_15,Baranov_16,Gadway_17,Birman_18}. These systems 
can form self-assembled lattice structures or can be loaded into external potentials 
imitating crystalline matter. The characteristics of these dipolar materials can be 
varied in a very wide range. Therefore the realization of the method considered in 
the paper looks feasible.

%Figure 1
\begin{figure*}[ht]
\centerline{
\includegraphics[width=8cm]{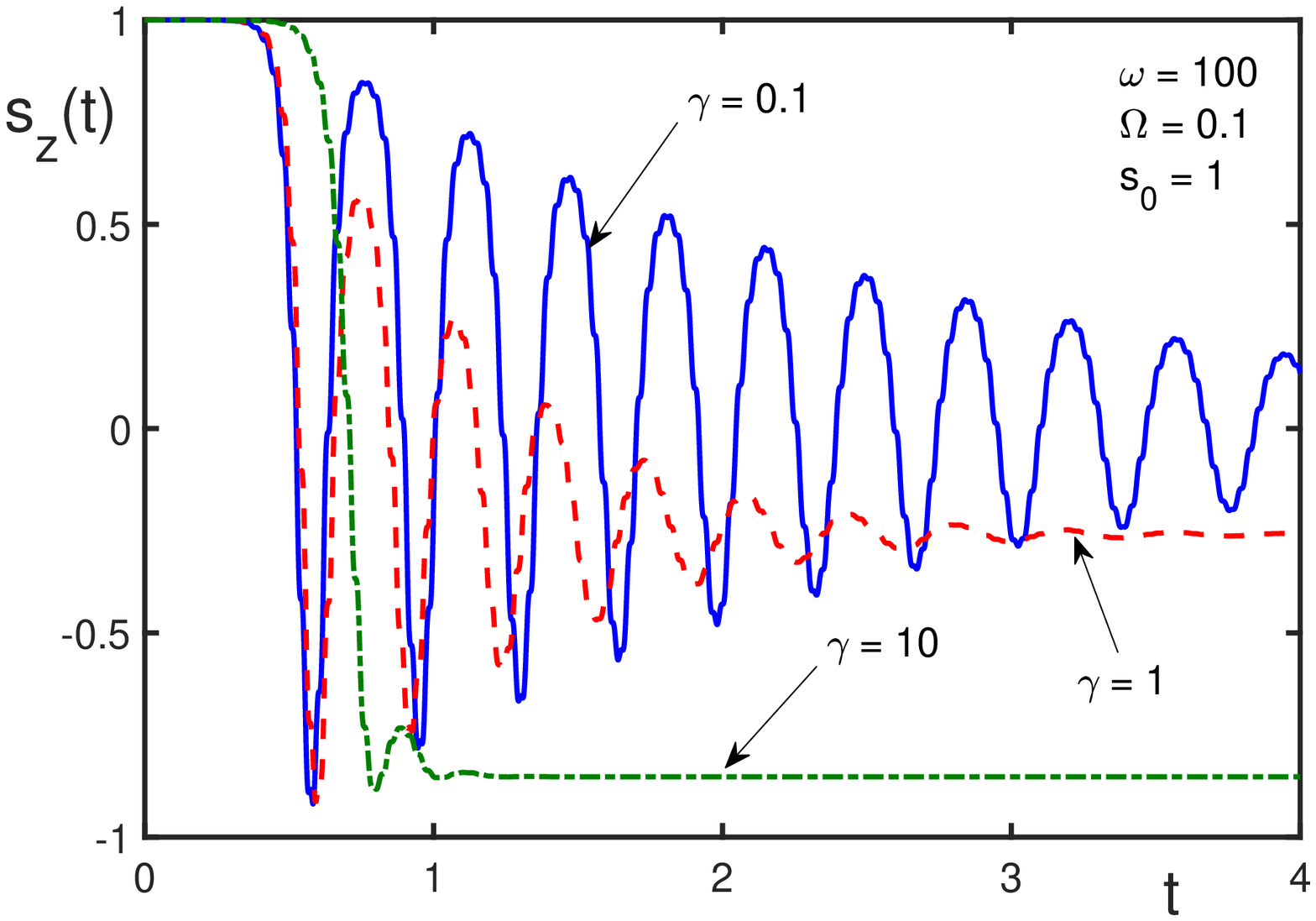} }
\vskip 3mm
\caption{Polarization switching as a function of time for different values of the 
resonator attenuation: $\gm=0.1$ (solid line); $\gm=1$ (dashed line); and $\gm=10$ 
(dashed-dotted line). Other parameters are: $\Omega=0.1$, $\om=100$, and $s_0=1$. 
Here and in the following figures, time is in units of $1/\gamma_2$ and all frequencies
are in units of $\gamma_2$. }
\label{fig:Fig.1}
\end{figure*}

%Figure 2
\begin{figure*}[ht]
\centerline{
\includegraphics[width=8cm]{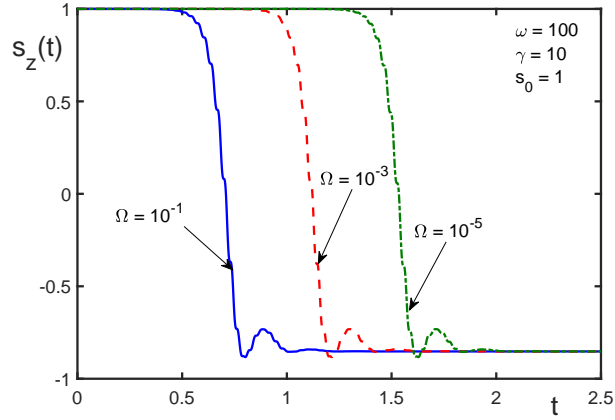} }
\vskip 3mm
\caption{Polarization switching for varying tunneling frequency: $\Omega = 10^{-5}$
(dashed-dotted line); $\Omega = 10^{-3}$ (dashed line); and $\Omega = 10^{-1}$ (solid 
line). Other parameters are: $\omega = 100$, $\gamma = 10$, and $s_0 = 1$.}
\label{fig:Fig.2}
\end{figure*}

%Figure 3
\begin{figure*}[ht]
\centerline{
\includegraphics[width=8cm]{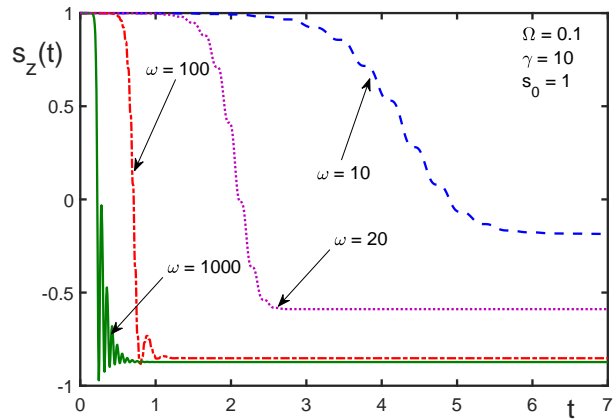} }
\vskip 3mm
\caption{Dependence of the polarization switching on the frequency: $\omega = 10$ 
(dashed line); $\omega = 20$ (dotted line); $\omega = 100$ (dashed-dotted line); and 
$\omega = 1000$ (solid line). Other parameters are: $\Omega = 0.1$, $\gamma = 10$, and
$s_0 = 1$.}
\label{fig:Fig.3}
\end{figure*}

%Figure 4
\begin{figure*}[ht]
\centerline{
\includegraphics[width=8cm]{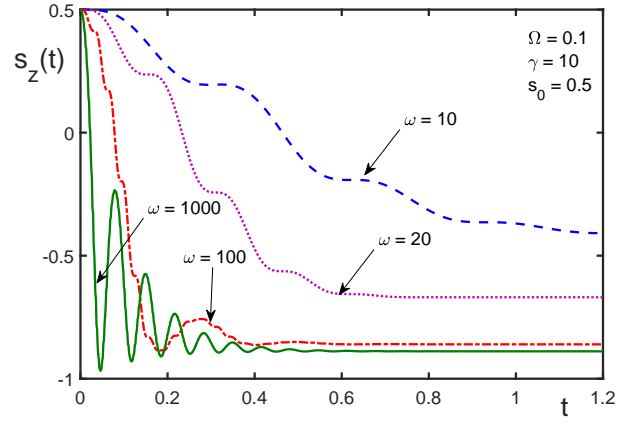} }
\vskip 3mm
\caption{Dependence of the polarization switching on the varying frequency $\om$ for
the same parameters as in the previous figure, $\Om=0.1$, $\gm=10$, but for the 
initial polarization $s_0 = 0.5$. Here: $\om=10$ (dashed line); $\om=20$ (dotted line); 
$\om=100$ (dashed-dotted line); and $\om=1000$ (solid line).}
\label{fig:Fig.4}
\end{figure*}

%Figure 5
\begin{figure*}[ht]
\centerline{
\includegraphics[width=8cm]{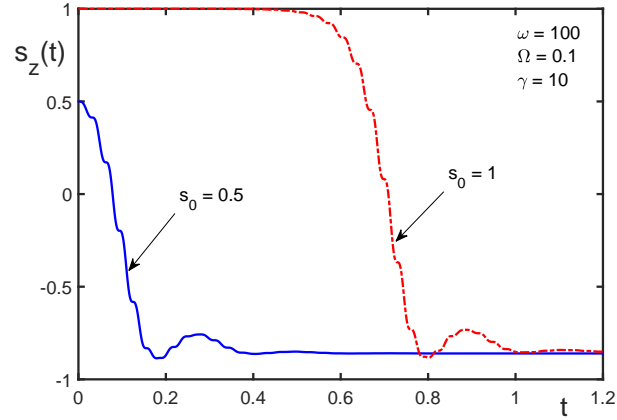} }
\vskip 3mm
\caption{Polarization switching for different initial polarizations: $s_0=0.5$ (solid line)
and $s_0=1$ (dashed-dotted line). Other parameters are: $\om=100$, $\Om=0.1$, and $\gm=10$.
For $s_0 < 1$, the delay time is practically absent. }
\label{fig:Fig.5}
\end{figure*}

%\Figure 6
\begin{figure*}[ht]
\centerline{\hbox{
\includegraphics[width=8cm]{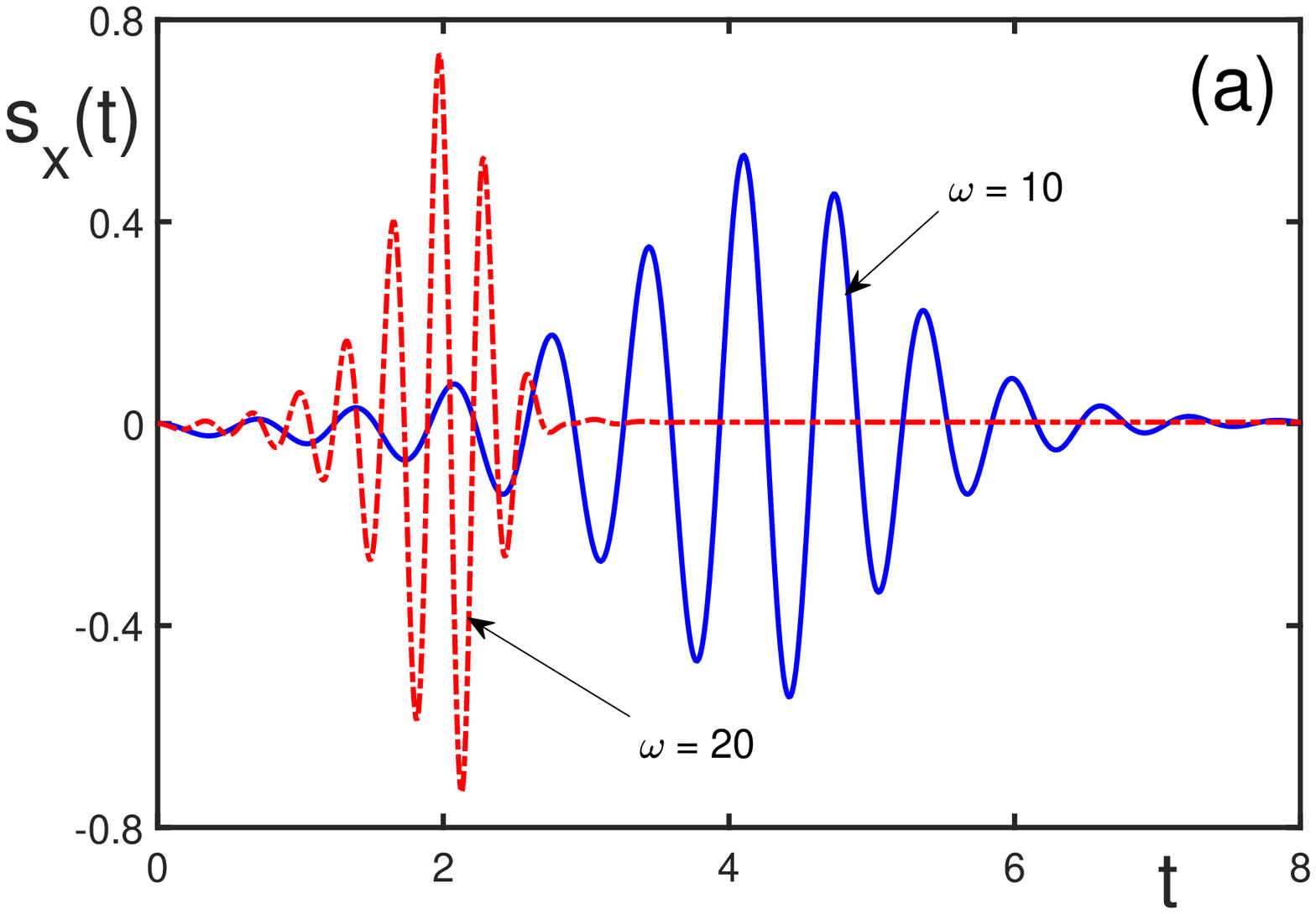} \hspace{1cm}
\includegraphics[width=8cm]{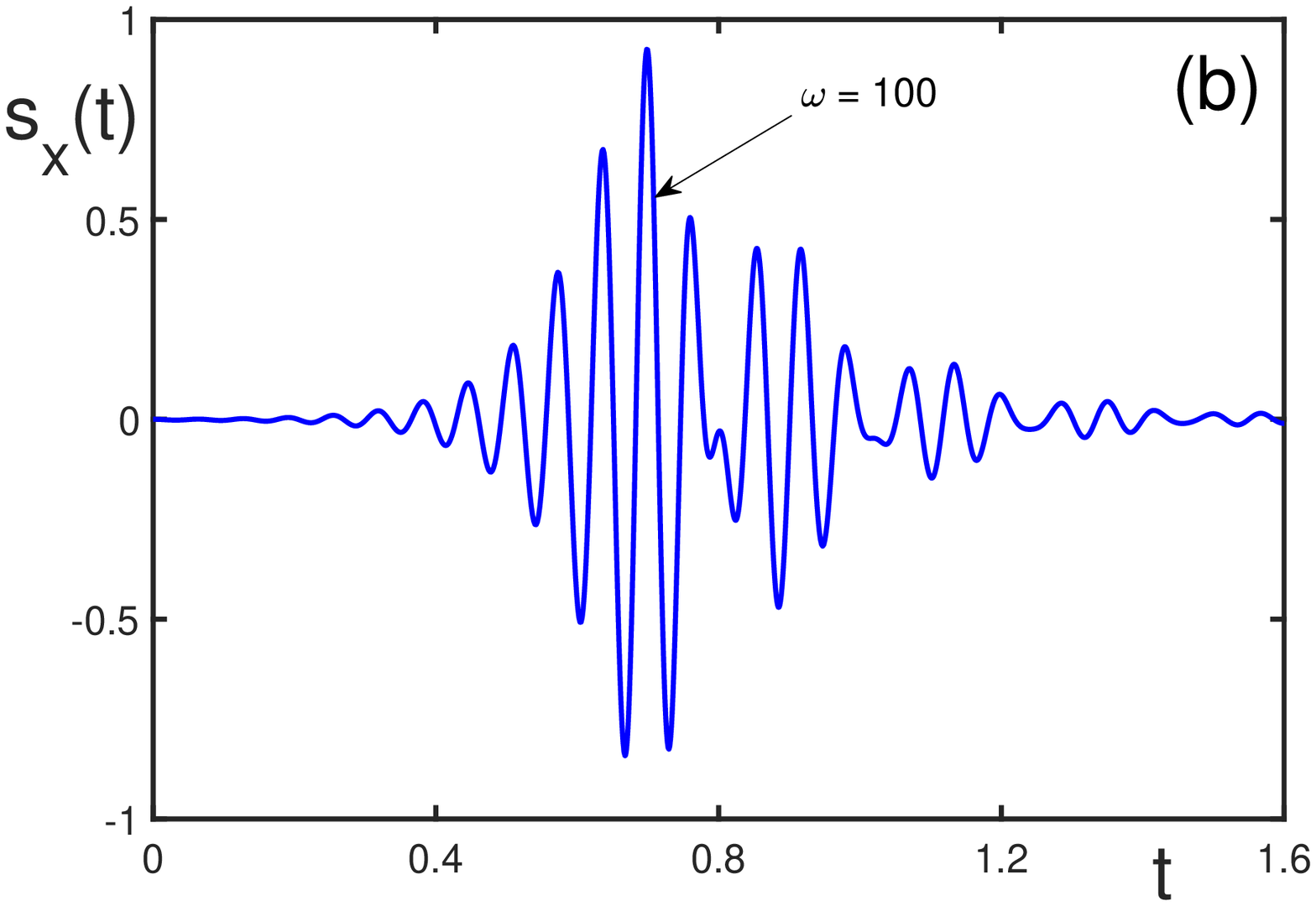} }   }
\caption{Temporal dependence of the transverse polarization component $s_x$, with the
parameters $\Om=0.1$, $\gm=10$, and $s_0=1$, for different frequencies: (a) $\om=10$ 
(solid line); $\om=20$ (dashed-dotted line); (b) $\om=100$ (solid line).   
}
\label{fig:Fig.6}
\end{figure*}

%\Figure 7
\begin{figure*}[ht]
\centerline{\hbox{
\includegraphics[width=8cm]{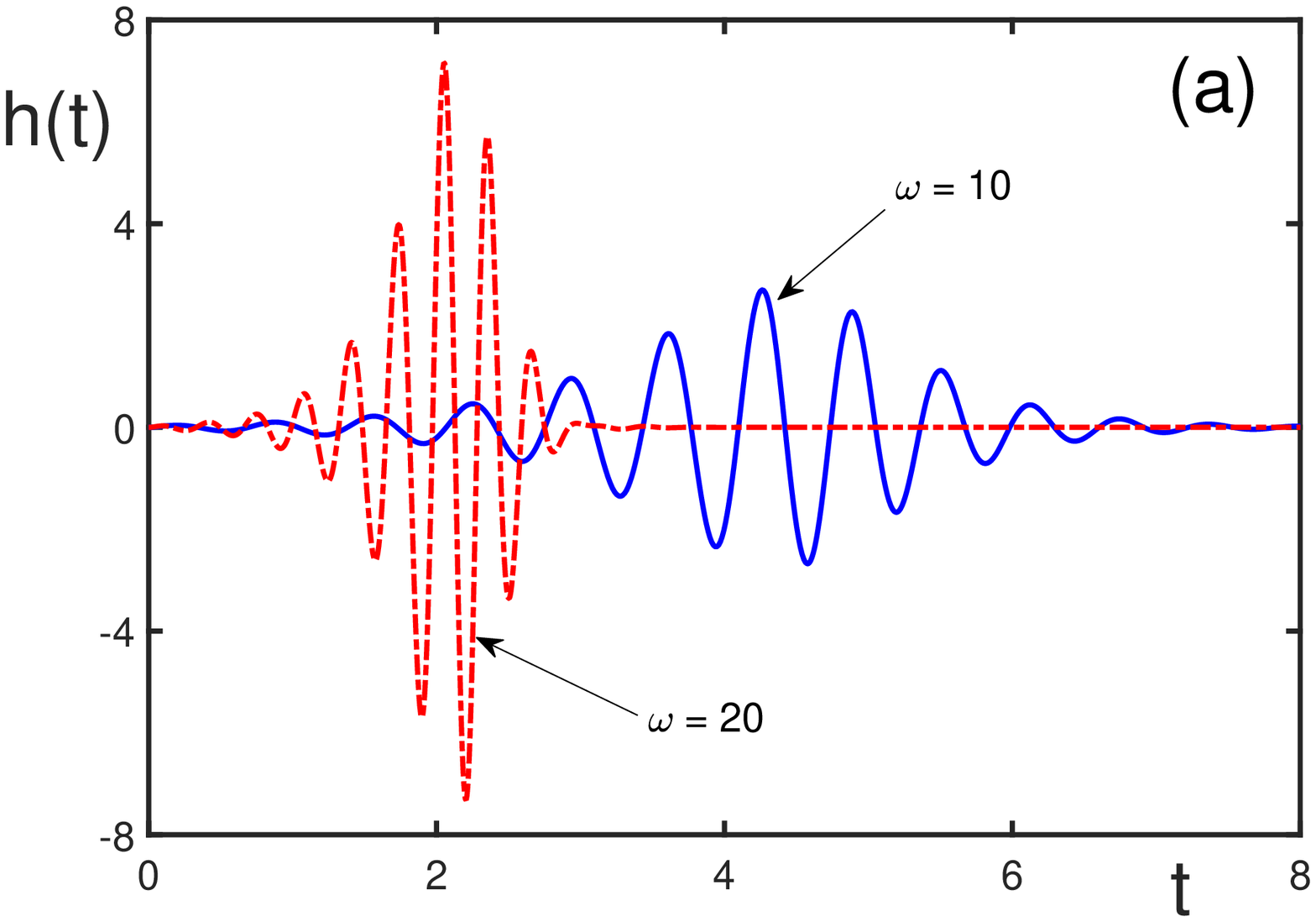} \hspace{1cm}
\includegraphics[width=8cm]{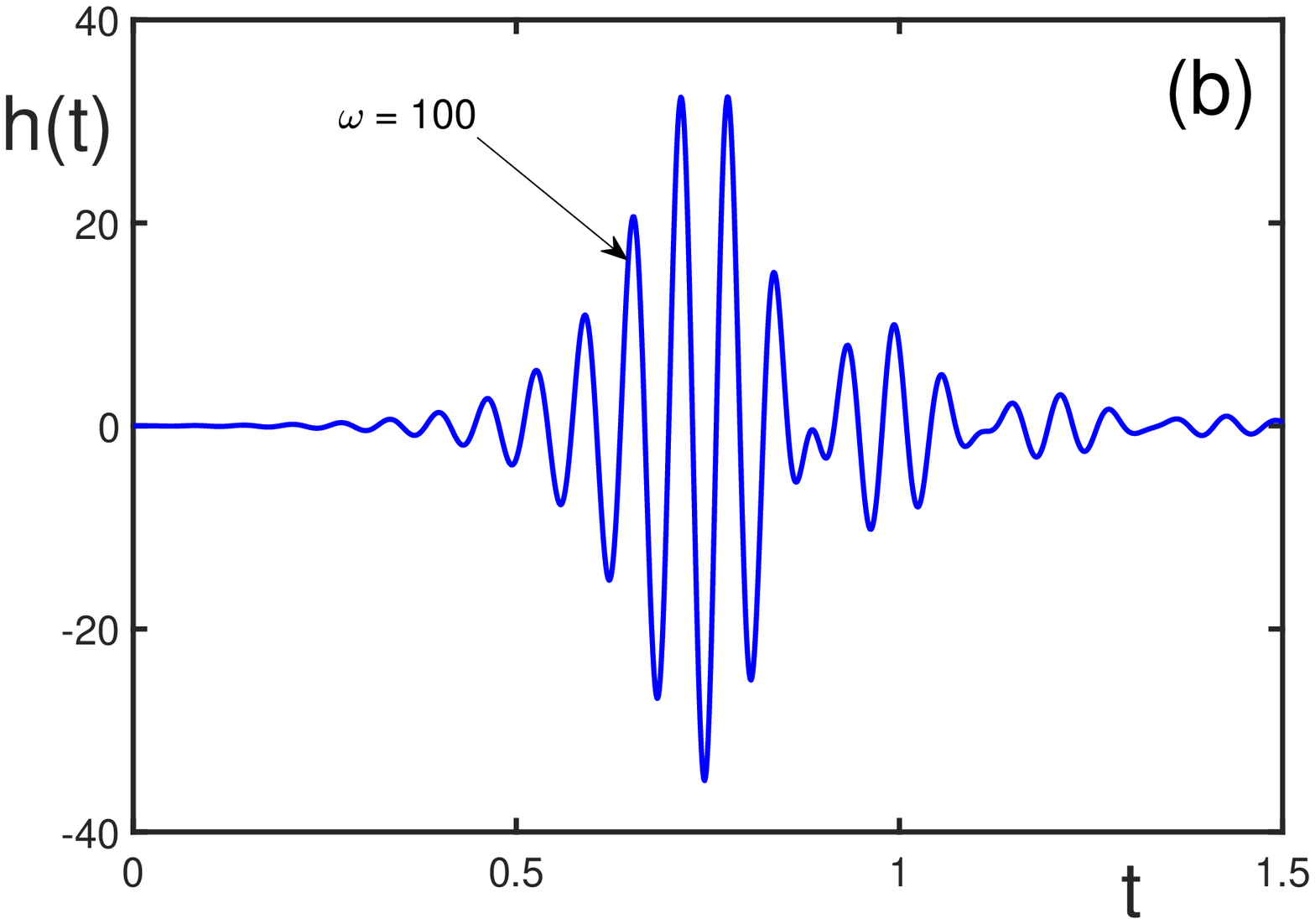} }   }
\caption{Dimensionless electric field inside the cavity $h(t)$, as a function of 
time with the parameters $\Om=0.1$, $\gm=10$, and $s_0=1$, for different frequencies: 
(a) $\om=10$ (solid line); $\om=20$ (dashed-dotted line); (b) $\om=100$ (solid line).   
}
\label{fig:Fig.7}
\end{figure*}

\end{document}